# Protein contact map prediction using bi-directional recurrent neural network

Yuhong Wang*[1], Wei Li[2], Hongmao Sun [1], Kennie Cruz-Gutierrez [1]

[1] National Center for Advancing Translational Sciences, 9800 Medical Center Drive, Rockville, Maryland 20875, USA

[2] School of life science, Jilin University, 2699 Qianjin Street, Changchun 130012, P. R. China

**Abstract**:

Given native 2D contact map, protein 3D structure could be reconstructed with accuracy of 2Å or better, and such reconstruction is a feasible computational approach for protein folding problem. The prediction accuracy from traditional methods is generally too poor to useful, but the recent deep learning model has significantly improved the accuracy. In this study, we proposed a neural network model comprising a bi-directional recurrent neural network and artificial neural network. Over the non-redundant database of all available protein 3D structures in Protein Data Bank, this deep learning model achieved an accuracy of 0.80, much higher than those of previous models. This study represents a major breakthrough in protein 2D contact map prediction and likely a major step forward for the protein folding problem.

**Introduction**

Reconstruction of protein 3D structure from 2D contact map is a computational approach for protein folding problem (Anfinsen 1973). Given native contact map, protein 3D structure could be reconstructed with accuracy of 2Å or better (Chen and Shen 2012) (Vassura et al. 2008).

Roughly there are three different groups of methods for protein contact map prediction: evolutionary coupling analysis (ECA), supervised machine learning (ML), and deep learning (DL). ECA depends upon protein sequence homology and predicts contacts by utilizing co-evolved residues in proteins, and it includes EVfold (Marks et al. 2011) , PSICOV (Jones et al. 2012), CCMpred (Seemayer, Gruber, and Söding 2014), Gremlin (Kamisetty, Ovchinnikov, and Baker 2013), DCA and others (Ekeberg et al. 2013) (Göbel et al. 1994) (Morcos et al. 2011). Some examples of ML include SVMSEQ (Wu and Zhang 2008) , CMAPpro (Di Lena, Nagata, and Baldi 2012), PconsC2 (Skwark et al. 2014), MetaPSICOV (Jones et al. 2015), PhyCMAP (Z. Wang and Xu 2013), and CoinDCA-NN (Ma et al. 2015). Unfortunately, these methods generally have very poor performance; the accuracy is around 0.20-0.30. The recently developed DL method (S. Wang et al. 2017) significantly improved the prediction accuracy.

In our previous study (Y. Wang et al. 2017), we showed that the intra-protein binding peptide fragments have specific and intrinsic sequence patterns, distinct from non-binding peptide fragments, and they can be predicted with an accuracy of up to 90%. Basing upon these findings, in this study, we use bidirectional recurrent neural network (biRNN), one common and effective variant of RNN, for protein contact map prediction. RNN is designed for sequence data analysis, it is able to capture long range dependence and relationship, and it has been successfully applied in problems from speech recognition, machine translation, text data analysis, to time series prediction.

**Results**

From 12,946 protein X-ray structures in the precompiled culled PDB list (G. Wang and Dunbrack 2003) from Protein Data Bank (PDB) (Berman et al. 2000), we used 8,646 proteins of size from 40 to 300 to extract contact and non-contact samples. The goal of the precompiled culled PDB list is to create a non-redundant coverage for all available protein structures. Proteins in this list have an amino acid percent identify < 50%, a resolution better than 2.0Å, and a R-factor smaller than 2.5.

8,646 proteins are randomly split into training data set of 8,056 proteins, validation data set of 283 proteins, and test data set of 307 proteins to generate, respectively, 1,746,885, 65,826, and 61,317 samples of 1:1 ratio for contact and non-contact samples.

Using the designed neural network, a biRNN followed by a fully connected neural network (fcNN), we optimized the hyper parameters by grid search (Table 1). Using 2048 hidden states in the RNN model, 2048 neurons in the fcNN, 0.00001 for the regularization coefficient, and 0.00005 for the initial learning rate, the prediction accuracy, AUC-ROC, false positive rate, and false negative rate are 0.80, 0.886, 0.158, and 0.240, respectively, on the test data set of 61,317 samples. The AUC-ROC curve is plotted in Figure 2. The loss of training data set is plotted versus iterations in Figure 3, and the prediction accuracy of validation data set shows similar profiles (Figure 4). This accuracy of 0.80 is based upon the default cutoff of 0.5 for contact and non-contact prediction, and it is increased to 0.815 if the optimal cutoff from AUC-ROC is used.

As a negative control, we randomly assigned labels of contact and non-contact for all samples and performed the training using the identical neural network model and hyper parameters. The accuracy, AUC-ROC, false positive rate and false negative rate on the test data set are 0.503, 0.501, 0.025, and 0.975, respectively. These numbers are expected for random samples.

Generally it is much more difficult to predict contact for two positions of great sequence distance; such contacts are determined by long distance and poorly understood interactions. We separate the test data set into subsets of sequence distance > 16, 32, 64, and 128 residues (Table 2). Apparently, the accuracy degrades as sequence distance increases; however even at a sequence distance of 128, the accuracy remains decent at 0.743.

**Discussion**

Over the data set of 8,646 proteins, our designed neural network model is able to predict protein contact map at an accuracy of 0.80. Even at a sequence distance of > 128, the accuracy remains decent at 0.743.

We compared our results with previous studies, in particular with those from the recent ultra deep neural network. In this study, we did not separate proteins into membrane and non-membrane ones, and our results are only comparable to the results in Table 2 and 3 (S. Wang et al. 2017). Also in this study, we excluded those samples with sequence distance < 10, so our results are comparable with the samples of medium and long ranges in Table 2 and 3. Furthermore our accuracy is calculated over all contact samples, not just the top L (protein length) contacts, our results are more stringent than those for top L contacts.

Since the models before the recent ultra-deep neural network only achieved accuracy, even over much smaller data set, around 0.20 (S. Wang et al. 2017), we will not discuss them further. For the ultra deep neural network, the best comparable accuracy is 0.55. Interestingly, the accuracy numbers seem to become worse as sequence distance decreases; this is opposite to our findings in this study.

Two factors likely contributed to the much improved prediction accuracy for protein contact map prediction in this study. First, RNN is designed for sequence data analysis and capable in capturing long range dependency and interactions; protein sequence is just one kind of sequence data, and contacts, in particular those at great sequence distance, are determined by long range interactions. Generally speaking, traditional perceptrons and convolution neural network (CNN) are less efficient than RNN for modeling sequence data.

Second, in this study, we used a more strict criterion in deciding whether two residues are in contact or not; the minimum distance between all heavy atoms of two residues has to be < 4.5 Å. In the ultra deep neural network, they used a cutoff of 8.0 Å between C beta atoms. The criteria used in earlier researches are even loose. The sizes of amino acids' side chains differ substantially, and such loose criterion could lead to high false positive and negative rates. We tested this by using the same cutoff as in the ultra deep neural network model. Using the identical neural network model and hyper parameters, the accuracy and AUC-ROC are 0.754 and 0.839, respectively, much lower than those from using the strict cutoff.

Use of validation data set and regularization is a common method for avoiding over-fitting. In this study, regularization does not seem to have great impact on the prediction performance unless a high penalty is used; the use of validation data set is very effective in avoiding over-fitting (Table 1). Even though we used 2048 hidden states for RNN, the performance remains good even with much fewer hidden states. For example, with 128 hidden states and 128 neurons in the fcNN, the accuracy is 0.773, about 0.03 below the optimal 0.80.

As described above, the accuracy degrades as sequence distance increases; however, even at a sequence distance of 128, the accuracy remains decent at 0.743. While this shows that RNN is capable to model such long range interactions, it also suggests possibility of improvement such as adding attention mechanism (https://nlp.stanford.edu/pubs/emnlp15_attn.pdf).

Theoretically, 100% accurate prediction of protein contact map will essentially solve the protein folding problem (Anfinsen 1973). While this study is a major step forward, it alone may not be sufficient for high accuracy reconstruction of protein 3D structure. For typical single domain and globular proteins, only a small fraction of possible pairs of residues form a contact. A false positive rate of 0.15 from this study could produce too many false contacts and thus disrupt good reconstruction of protein 3D structures from contact map. We are currently exploring new models while testing a few algorithms for reconstructing protein 3D structures from contact map.

**Materials and Methods**

From 12,946 protein X-ray structures in the precompiled culled PDB list (G. Wang and Dunbrack 2003) from Protein Data Bank (PDB) (Berman et al. 2000), we used 8,646 proteins of size from 40 to 300 to extract contact and non-contact samples. There are two considerations for setting upper bound of protein size at 300. First, most of single domain globular proteins have a size below 300. Second, greater protein size requires more GPU memory and smaller training batch size, which are beyond our available resource.

Extraction of contact and non-contact samples

For two positions, $i$th and $j$th residues, on a protein sequence, it is considered in contact if the minimum spatial distance between all heavy atoms of the two residues is smaller than 4.5 Å and in non-contact if the minimum distance is greater than 12 Å. Two positions with sequence distance smaller than 10 is excluded. For contact residues, a smaller cutoff (<4.5 Å) will produce slightly better prediction accuracy at the cost of reduced training data size. As seen in this and previous study, 4.5 Å is a good compromise. For non-contact samples, 12 Å is sufficiently large; no meaningful interactions, except Coulombic, between two heavy atoms could exist at this distance. In this study, contact samples are duplicated three times in order to increase sample size and keep the samples balanced, in other words having comparable numbers of contact and non-contact samples. A total of 1,874,064 samples were extracted with half of contact and half of non-contact samples.

Design of the neural network

Deep Learning (LeCun, Bengio, and Hinton 2015) methods, as representation learning methods, allow deep neural networks discovering the representations from raw data for specific tasks such as classification and detection. Supervised learning is the most common form of machine learning which deep learning improves the state-of-the-art of most supervised learning problems. With the help of the ground truth or label of data set, deep learning can learn better representation to predict such ground truth. A loss function captures the distance between the current output of the neural network and the ground truth, then the network propagates the error backwards to adjust all the parameters (weights) in the neural network. In this way, the loss or distance can be significantly reduced after the training process. The classification of contact and non-contact between two residues on a protein is supervised learning with the ground truth as if the two residues are in contact or non-contact. Thus, we use deep learning to learn better features and get better classification performance.

Protein sequence is obviously one kind of sequence data. For sequence data, recurrent neural network (RNN) is a natural architect and more efficient than artificial neural network (ANN). RNN has loops, which allow information to be passed from one step of the network to the next. In the past decade, RNN has incredible success in various problems such as speech recognition, language modeling, and translation, and one key to the successes is the use of Long Term Short Term Memory (LSTM) model, a special kind of RNN.

LSTM networks were introduced by Hochreiter & Schmidhuber (Hochreiter, Schmidhuber, and Jurgen 1997). The key to LSTMs is the cell state. Cell state is like a conveyor belt, which runs through a sequence, with linear interactions with each element in the sequence. Because of this unique design, LSTM allows information to flow along a sequence unchanged and thus is capable of modeling and capturing distant relationships.

In bi-directional RNNs, two independent RNNs are placed together in opposite direction. The input sequence is fed in normal time order for one network, and in reverse time order for the other. The outputs of the two networks are concatenated at each time step. This structure allows the networks to have both backward and forward information about the sequence at every time step (https://www.intel.ai/wp-content/uploads/sites/53/2017/06/BRNN.pdf)

We designed a composite neural network for prediction of protein contact map. This network starts with a bi-directional RNN, followed by a fully connected neural network of one hidden layer, and one output layer for contact and non-contact. The input to the bi-directional RNN is the entire protein sequence with each residue encoded by a one-hot vector. The $i$th and $j$th outputs of the forward RNN and the corresponding outputs of the reverse RNN are concatenated to form the input for the fully connected neural network. The hidden layers use the activation function of Rectified Linear Units (ReLU) (Nair, Vinod, and Hinton 2010) which can introduce nonlinearity into the presentation learning. After the hidden layers, Softmax layer is used as the classification layer (or the output layer of two nodes for contact or non-contact). Backpropagation is used for training the network (Rumelhart et al. 1986).

Optimization of the loss function is carried out by mini-batch of a size 256 and the ADAM optimizer (6), which is implemented as tf.train.AdamOptimizer in the Tensorflow library (www.tensorflow.org). The regularization coefficient and starting learning rate were optimized after a grid search (Table 2).

The neural network training and prediction were performed on a Dell PowerEdge R940xa server with four Intel Xeon Platinum 8160 processors (each with 24 cores), 3TB of RAM and four 16GB NVIDIA Tesla V100 graphic processing unit, installed with Ubuntu 16.04.6 distribution, python 3.5, CUDA driver version 10.0, cuDNN version 7.4, TensorRT 5.1 and Tensorflow 1.13.1. The python program was written to implement the neural network model (Figure 1) and optimize the loss function.

Contact and non-contact samples were randomly split into three data sets: 80% for training, 10% for validation, and 10% for testing. Out of 8,646 proteins, 283 are used for validation and 307 for testing. The training process was constantly monitored by checking the accuracy of the validation data set, and it was terminated when the performance was noticeably worsened. It usually took up to 20 epochs and 2-4 days. The trained models were applied to the test data set for benchmarking.

For negative control, the label of contact and non-contact samples was randomly assigned as contact (1) or non-contact (0), and the same training procedure and benchmarking were performed.

**Figures**

Figure 1. Illustration of bi-directional recurrent neural network

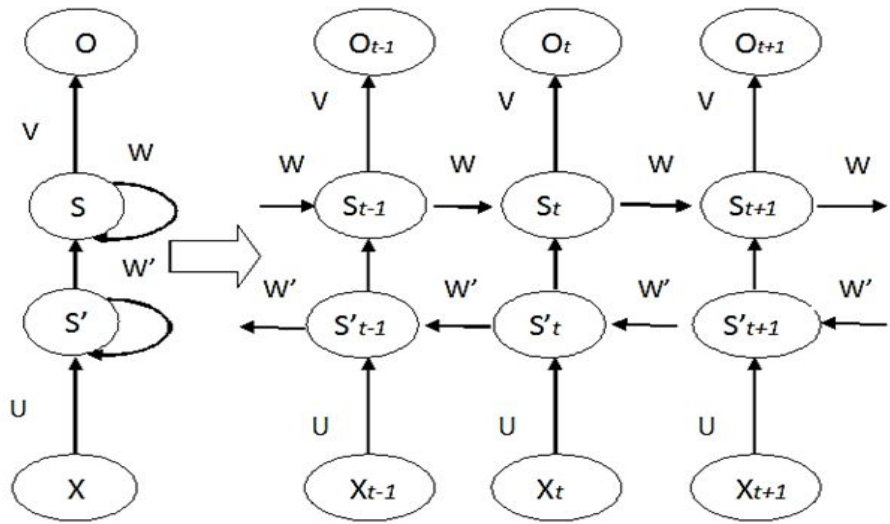

Figure 2. AUC-ROC curves for contact and non-contact samples of the test data set. Total sample size of test data is 61317, and the AUC-ROC are 0.886.

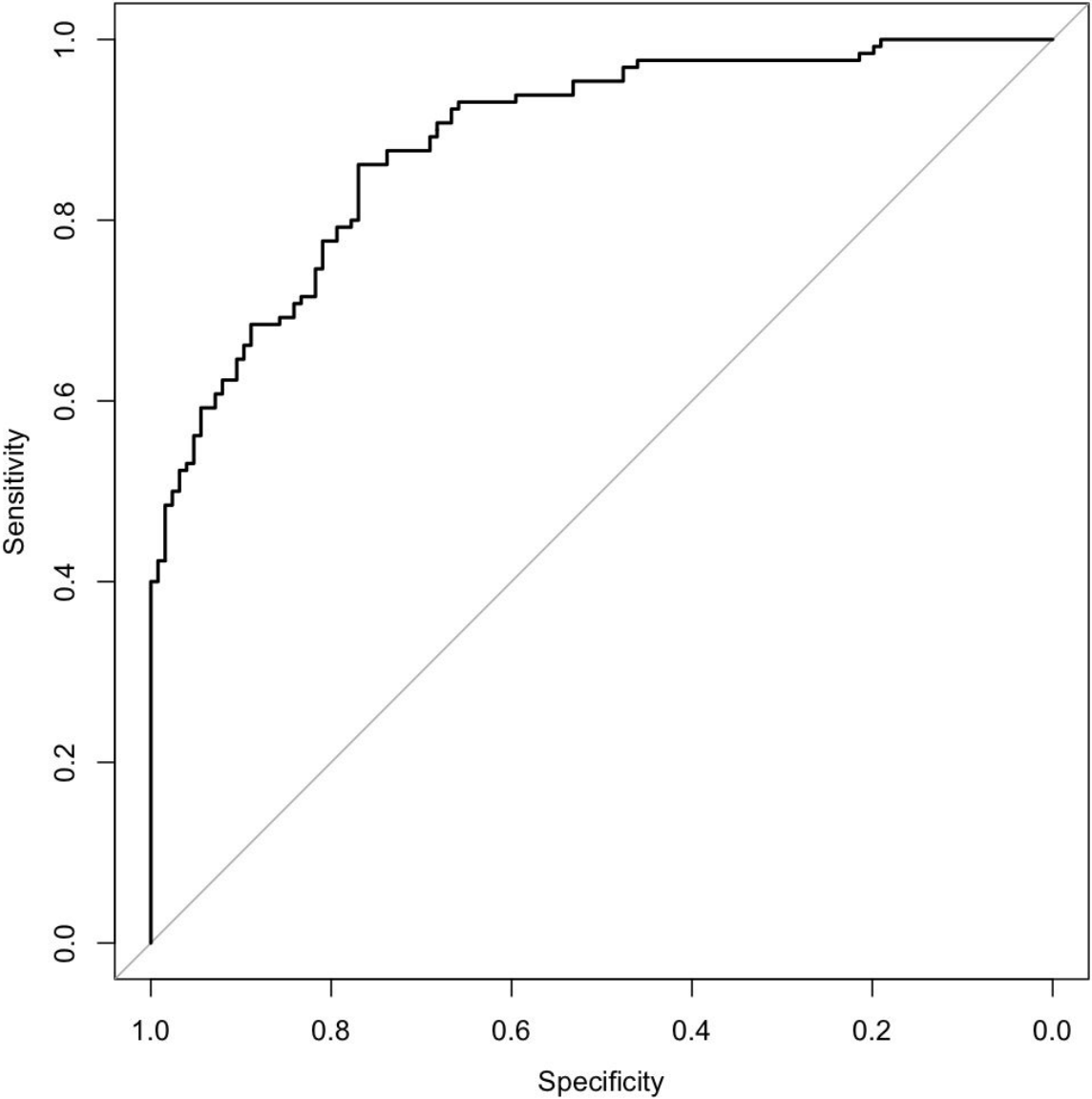

Figure 3. Model loss on training data sets for contact and non-contact samples. Sample size of the training data set is 1,746,885.

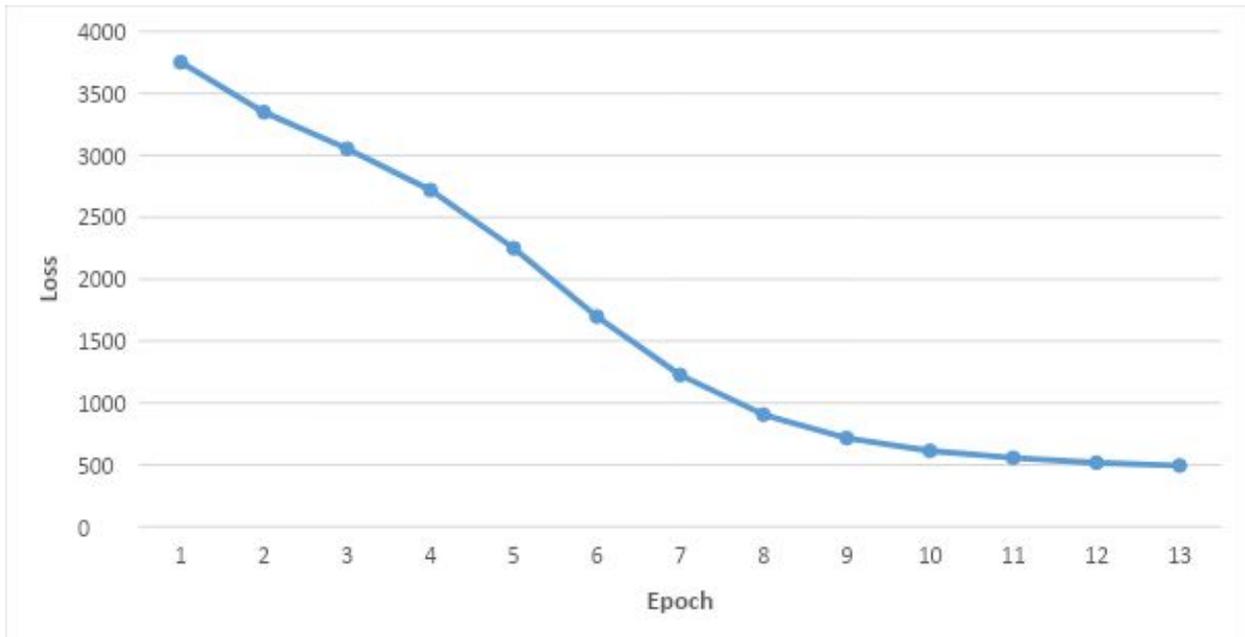

Figure 4. Model accuracy on validation data sets for contact and non-contact samples. Sample size of the validation data set is 65862.

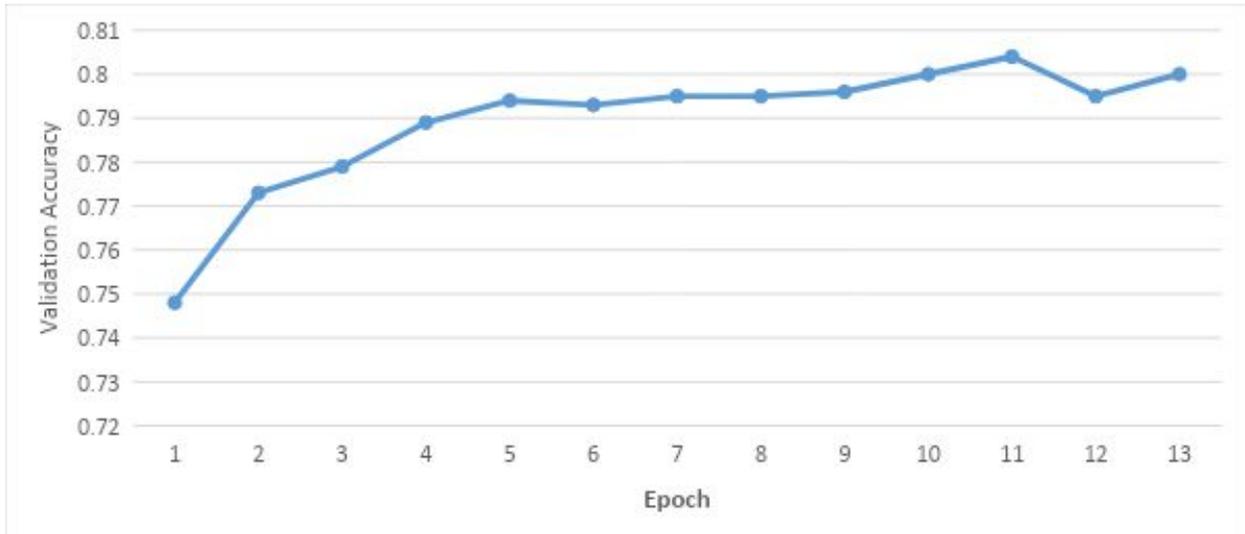

**Tables**

Table 1. Optimization of hyper parameters for the designed neural network.

| Reg Coef[1] | Learning Rate | Hidden State[2] | Hidden Layer[3] | Training accuracy | Validation accuracy | Test accuracy | Test AUC-ROC | False positive rate | False negative rate |
|---|---|---|---|---|---|---|---|---|---|
| 0.00000 | 0.00005 | 1024 | 1024 | 0.911 | 0.788 | 0.792 | 0.875 | 0.17597 | 0.24101 |
| 0.00001 | 0.00005 | 1024 | 1024 | 0.988 | 0.792 | 0.792 | 0.871 | 0.18523 | 0.23131 |
| 0.00010 | 0.00005 | 1024 | 1024 | 0.886 | 0.784 | 0.787 | 0.874 | 0.20974 | 0.21502 |
| 0.00100 | 0.00005 | 1024 | 1024 | 0.766 | 0.758 | 0.770 | 0.854 | 0.16366 | 0.29656 |
| | | | | | | | | | |
| 0.00010 | 0.00005 | 128 | 128 | 0.798 | 0.759 | 0.773 | 0.857 | 0.24590 | 0.20787 |
| 0.00010 | 0.00005 | 256 | 256 | 0.855 | 0.770 | 0.770 | 0.854 | 0.21552 | 0.24440 |
| 0.00010 | 0.00005 | 512 | 512 | 0.880 | 0.778 | 0.783 | 0.868 | 0.22541 | 0.20883 |
| 0.00010 | 0.00005 | 1024 | 1024 | 0.886 | 0.784 | 0.787 | 0.874 | 0.20974 | 0.21502 |
| 0.00010 | 0.00005 | 2048 | 2048 | 0.858 | 0.787 | 0.789 | 0.875 | 0.17108 | 0.24975 |
| | | | | | | | | | |
| 0.00001 | 0.00005 | 2048 | 2048 | 0.986 | 0.800 | 0.800 | 0.886 | 0.15867 | 0.24075 |
| 0.00010 | 0.00005 | 2048 | 2048 | 0.858 | 0.787 | 0.789 | 0.875 | 0.17108 | 0.24975 |

[1] Regularization coefficient. [2] Number of hidden states in the RNN. [3] Number of nodes in the hidden layer.

Table 2. Prediction accuracy, false positive rate, and false negative rate at different sequence distances on the test data set.

| Sequence Distance (>) | Number of samples | Accuracy | False positive rate | False negative rate |
|---|---|---|---|---|
| 16 | 53846 | 0.800 | 0.188 | 0.209 |
| 32 | 37887 | 0.795 | 0.190 | 0.238 |
| 64 | 18868 | 0.757 | 0.236 | 0.249 |
| 128 | 6014 | 0.743 | 0.244 | 0.265 |

**Preferences**

Anfinsen, C. B. 1973. "Principles That Govern the Folding of Protein Chains." *Science* 181 (4096): 223–30.
Berman, H. M., J. Westbrook, Z. Feng, G. Gilliland, T. N. Bhat, H. Weissig, I. N. Shindyalov, and P. E. Bourne. 2000. "The Protein Data Bank." *Nucleic Acids Research* 28 (1): 235–42.
Chen, Jun, and Hong-Bin Shen. 2012. "Glocal: Reconstructing Protein 3D Structure from 2D Contact Map by Combining Global and Local Optimization Schemes." *Current Bioinformatics*. https://doi.org/10.2174/157489312800604381.
Di Lena, Pietro, Ken Nagata, and Pierre Baldi. 2012. "Deep Architectures for Protein Contact Map Prediction." *Bioinformatics* 28 (19): 2449–57.
Ekeberg, Magnus, Cecilia Lövkvist, Yueheng Lan, Martin Weigt, and Erik Aurell. 2013. "Improved Contact Prediction in Proteins: Using Pseudolikelihoods to Infer Potts Models." *Physical Review. E, Statistical, Nonlinear, and Soft Matter Physics* 87 (1): 012707.
Göbel, U., C. Sander, R. Schneider, and A. Valencia. 1994. "Correlated Mutations and Residue Contacts in Proteins." *Proteins* 18 (4): 309–17.
Hochreiter, Sepp; And Schmidhuber, and Jurgen. 1997. "Long Short-Term Memory." *Neural Computation* 9 (8): 45.
Jones, David T., Daniel W. A. Buchan, Domenico Cozzetto, and Massimiliano Pontil. 2012. "PSICOV: Precise Structural Contact Prediction Using Sparse Inverse Covariance Estimation on Large Multiple Sequence Alignments." *Bioinformatics* 28 (2): 184–90.
Jones, David T., Tanya Singh, Tomasz Kosciolek, and Stuart Tetchner. 2015. "MetaPSICOV: Combining Coevolution Methods for Accurate Prediction of Contacts and Long Range Hydrogen Bonding in Proteins." *Bioinformatics* 31 (7): 999–1006.
Kamisetty, Hetunandan, Sergey Ovchinnikov, and David Baker. 2013. "Assessing the Utility of Coevolution-Based Residue–residue Contact Predictions in a Sequence- and Structure-Rich Era." *Proceedings of the National Academy of Sciences of the United States of America* 110 (39): 15674–79.
LeCun, Y., Y. Bengio, and G. Hinton. 2015. "Deep Learning." *Nature* 521 (7553): 436–44.
Ma, Jianzhu, Sheng Wang, Zhiyong Wang, and Jinbo Xu. 2015. "Protein Contact Prediction by Integrating Joint Evolutionary Coupling Analysis and Supervised Learning." *Bioinformatics* 31 (21): 3506–13.
Marks, Debora S., Lucy J. Colwell, Robert Sheridan, Thomas A. Hopf, Andrea Pagnani, Riccardo Zecchina, and Chris Sander. 2011. "Protein 3D Structure Computed from Evolutionary Sequence Variation." *PloS One* 6 (12): e28766.
Morcos, F., A. Pagnani, B. Lunt, A. Bertolino, D. S. Marks, C. Sander, R. Zecchina, J. N. Onuchic, T. Hwa, and M. Weigt. 2011. "Direct-Coupling Analysis of Residue Coevolution Captures Native Contacts across Many Protein Families." *Proceedings of the National Academy of Sciences*. https://doi.org/10.1073/pnas.1111471108.
Nair, Vinod, and And Geoffrey E. Hinton. 2010. "Rectified Linear Units Improve Restricted Boltzmann Machines." *Proceedings of the 27th International Conference on Machine Learning (ICML-10)*.

Rumelhart, David E. ;. Hinton, Geoffrey E. ;. Williams, and J. Ronald. 1986. "Learning Representation by Back-Propagating Errors." *Nature* 323 (6088): 4.

Seemayer, Stefan, Markus Gruber, and Johannes Söding. 2014. "CCMpred--Fast and Precise Prediction of Protein Residue-Residue Contacts from Correlated Mutations." *Bioinformatics* 30 (21): 3128–30.

Skwark, Marcin J., Daniele Raimondi, Mirco Michel, and Arne Elofsson. 2014. "Improved Contact Predictions Using the Recognition of Protein like Contact Patterns." *PLoS Computational Biology* 10 (11): e1003889.

Vassura, Marco, Luciano Margara, Pietro Di Lena, Filippo Medri, Piero Fariselli, and Rita Casadio. 2008. "Reconstruction of 3D Structures from Protein Contact Maps." *IEEE/ACM Transactions on Computational Biology and Bioinformatics / IEEE, ACM* 5 (3): 357–67.

Wang, G., and R. L. Dunbrack. 2003. "PISCES: A Protein Sequence Culling Server." *Bioinformatics* 19 (12): 1589–91.

Wang, Sheng, Siqi Sun, Zhen Li, Renyu Zhang, and Jinbo Xu. 2017. "Accurate De Novo Prediction of Protein Contact Map by Ultra-Deep Learning Model." *PLoS Computational Biology* 13 (1): e1005324.

Wang, Yuhong, Junzhou Huang, Wei Li, Sheng Wang, and Chuanfan Ding. 2017. "Specific and Intrinsic Sequence Patterns Extracted by Deep Learning from Intra-Protein Binding and Non-Binding Peptide Fragments." *Scientific Reports* 7 (1): 14916.

Wang, Zhiyong, and Jinbo Xu. 2013. "Predicting Protein Contact Map Using Evolutionary and Physical Constraints by Integer Programming." *Bioinformatics* 29 (13): i266–73.

Wu, Sitao, and Yang Zhang. 2008. "A Comprehensive Assessment of Sequence-Based and Template-Based Methods for Protein Contact Prediction." *Bioinformatics* 24 (7): 924–31.